\documentclass[final,5p,times,twocolumn]{elsarticle}

\usepackage{amssymb}
\usepackage{amsmath}
\usepackage{url}
\usepackage{upgreek}
\usepackage{subcaption}
\usepackage{xcolor}
\usepackage{soul}

\journal{Measurement}

\begin{document}

\begin{frontmatter}

\title{TPA-TCT analysis of the RD50-MPW4 monolithic pixel particle detector}

\author[1]{\mbox{Francisco Rogelio Palomo}\corref{cor}}
\ead{fpalomo@us.es}
\cortext[cor]{Corresponding author}

\author[1]{\mbox{Jorge Jiménez-Sánchez}}
\ead{jjsanchez@us.es}

\author[2]{\mbox{Moritz Wiehe}}
\ead{m.wiehe@cern.ch}

\author[3]{\mbox{Jory Sonneveld}} 
\ead{jory.sonneveld@cern.ch}

\author[4]{\mbox{Bernhard Pilsl}}
\ead{bernhard.pilsl@oeaw.ac.at}
      
\author[1]{\mbox{Fernando Muñoz-Chavero}}
\ead{fmunoz@us.es}

\author[5]{\mbox{Raimon Casanova}}
\ead{rcasanova@ifae.es}

\author[4]{\mbox{Christian Irmler}} 
\ead{christian.irmler@oeaw.ac.at}

\author[2]{\mbox{Patrick Sieberer}} 
\ead{patrick.sieberer@cern.ch}

\author[6]{\mbox{Chenfan Zhang}} 
\ead{chenfan@hep.ph.liv.ac.uk}

\author[7]{\mbox{Sinuo Zhang}} 
\ead{s.zhang@physik.uni-bonn.de}

\author[6]{\mbox{Eva Vilella}} 
\ead{vilella@hep.ph.liv.ac.uk}

\author[2]{\mbox{Michael Moll}} 
\ead{Michael.Moll@cern.ch}

\affiliation[1]{organization={Departamento de Ingeniería Electrónica, Universidad de Sevilla},
            city={Sevilla},
            country={Spain}}
            
\affiliation[2]{organization={Experimental Physics-Detector Technologies, CERN},
			city={Geneva},
			country={Switzerland}}

\affiliation[3]{organization={NIKHEF},
			city={Amsterdam},
			country={Netherlands}}

\affiliation[4]{organization={Marietta Blau Institute for Particle Physics of the Austrian Academy of Sciences},
			city={Vienna},
			country={Austria}}

\affiliation[5]{organization={Universidad Autónoma de Barcelona},
			city={Barcelona},
			country={Spain}}

\affiliation[6]{organization={Department of Physics, University of Liverpool},
			city={Liverpool},
			country={United Kingdom}}

\affiliation[7]{organization={University of Bonn},
			city={Bonn},
			country={Germany}}          
            
\begin{abstract}
The RD50-MPW4, a Depleted Monolithic Active Pixel Sensor (DMAPS) was analyzed using a Two Photon Absorption Transient Current Technique (TPA-TCT). This technique provides sensitivity maps with micrometer-scale spatial resolution, enabling the resolution of the boundaries of the detector's sensitive volume, even for small-area pixels (62 × 62 $\upmu$m$^2$ in this study). With a 3D resolution, the depletion depth, the boundaries of the detector electric field, the 3D hit detection efficiency and the charge sharing between neighboring pixels were measured. The RD50-MPW4, a multi-project wafer chip developed by the HV-CMOS working group within the CERN RD50 collaboration, features a 64 $\times$ 64 DMAPS pixel matrix. Illuminating the chip from the backside, the TPA-TCT technique can characterize any pixel element in the matrix because silicon is transparent for near infrared laser light (1550 nm). Electron-hole pairs are generated only around the light focal point, deep in the silicon, so that any charge collected is precisely only from the focal point. With the TPA-TCT technique, the RD50-MPW4 was found to be have a 100\% hit detection efficiency under specified conditions and an effective depletion depth of 226 $\upmu$m. It was also found that part of the charge in the periphery of the  pixel was collected in the neighboring pixel. A 3D map of the sensor clearly shows the in-pixel electronics and the limits of the depletion region.
\end{abstract}

\begin{keyword}
Two Photon Absorption (TPA) \sep 
Transient Current Technique (TCT) \sep 
Pulsed Femtosecond Laser \sep 
Monolithic Pixel Particle Detectors \sep 
Depleted Monolithic Active Pixel Sensor (DMAPS) \sep 
High-Voltage CMOS (HV-CMOS) \sep 
RD50-MPW4 chip
\end{keyword}

\end{frontmatter}

\section{Introduction}
Pixel detectors play a central role in high-energy physics (HEP), providing precise spatial measurements for charged particles and forming the core of modern tracking systems used in particle physics experiments \cite{Wermes-2018}. Although their primary use remains within high-energy physics, pixel detector technologies have found widespread application beyond fundamental research. Driven by the demanding requirements of HEP, these technologies have been successfully transferred to applied physics and other practical fields. Representative examples include their use in nuclear medicine imaging systems \cite{Kamtchou-2024}, in astrophysical instruments for the observation of high-energy cosmic phenomena \cite{Steinhebel-2024}, in radiation monitoring and safety within the nuclear energy sector \cite{Kim-2019}, and in industrial applications including non-destructive testing and radiation monitoring in harsh environments \cite{Jakubek-2009}.

Currently, the mainstream pixel detector for HEP (at the High Luminosity-Large Hadron Collider, HL-LHC) is a hybrid pixel detector \cite{Hartmann-2024} comprising a silicon sensor silicon bump bonded to a pixel readout electronics chip (see, for example \cite{RD53pixel-2025}). Alternatively, a monolithic CMOS pixel sensor for High Energy Physics (HEP) \cite{MonolithicCMOS-2023} integrates the sensor matrix and the readout electronics in the same chip. In comparison, hybrid arrangement is more expensive and resource-intensive to produce and has a higher material budget than the monolithic alternative. 

In the domain of monolithic pixel sensors, deployed MAPS (Monolithic Active Pixel Sensors) operate their sensor diodes in partial depletion mode; in high energy physics they were first deployed in STAR at RHIC at BNL \cite{StarMAPS-2008}, and ALICE was the first a large-scale experiment at the LHC at CERN \cite{Alice-2008} to use them in their Inner Tracking System, iteration 2 (ITS2)  with the ALICE Pixel Detector (ALPIDE) \cite{Alpide-2016,ITS2-2025,ITS2TDR-2014,LS2Upgrade-2024}. 

The next step are DMAPS (Depleted MAPS) sensors. Their sensors are designed to operate in a fully depleted mode. Their charge collection takes place through carrier drift in the strong electric field of the detector. That ensures a faster response than non-depleted MAPS, where carriers move essentially by diffusion. Full depletion of the pixel matrix means that the silicon substrate needs a high resistivity, in the 1-2 k$\Omega$$\cdot$cm range.

The RD50 collaboration, and more recently the DRD3 collaboration, developed a DMAPS sensor named the RD50-MPW4 \cite{DMAPS-2022} where MPW stands for Multi-Project Wafer. A working prototype of the RD50-MPW4 chip has existed since 2024 \cite{Vilella-2025}, and first studies of its performance and radiation tolerance under broad electron beam irradiation have been completed \cite{Pilsl-2024,Pilsl-2025}. At the moment, evolutions of the MPW4 are considered for the LHCb upgrade in the framework of the RadPix project \cite{RadPix-2026}.

As DMAPS arrays use reverse biased diodes as sensing elements, it is important to characterize their depletion depth, electric field boundaries, Hit Detection Efficiency (HDE) and Charge Sharing (CS). An appropriate technique is the TPA-TCT (Two Photon Absorption-Transient Current Technique), previously demonstrated for characterization of DMAPS detector diodes \cite{Palomo-2025}. 

TPA-TCT is a laser analysis technique to assess the quality of a sensor diode, capable of measuring the detector active volume thickness (depletion depth, depletion volume boundaries), to characterize HDE and response time, the CS between neighboring pixels, as well as qualitative information relative to the geometry of the electric field in the depletion volume in three dimensions. In TPA, the laser wavelength, so the photon energy, is below the silicon bandgap (\(\lambda\)>1150 nm), and consequently the linear absorption of light is negligible. In TPA, light is absorbed nonlinearly, i.e., only at sufficiently high intensities, such that carrier generation occurs predominantly near the focal point. In TCT, the transient signal generated by charge carriers as they move to the collecting electrodes is recorded and analyzed. As the laser photo-ionization is point generated, the technique can generate 3D sensitivity maps of a non-backside processed reverse biased detector diode.

In this work, we characterize a pixel of the High Voltage CMOS (HVCMOS)  detector RD50-MPW4 with the TPA-TCT technique. The uniformity of the sensor electric field as well as the depletion depth are useful measures to characterize a detector. The electric field uniformity is related to the charge collection efficiency (CCE) \cite{CCE-1990,CCE-2019}. Furthermore, the depletion depth gives an estimate of the detector junction capacitance \cite{Hartmann-2024} and effective space charge concentration \cite{EdgeTCT-2022}. TPA-TCT can be used to create a precise active-volume map, which will be presented here for the RD50-MPW4. This work has previously been carried out for an earlier prototype, the RD50-MPW2 \cite{Palomo-2025}. Compared to the RD50-MPW2, the RD50-MPW4 now has digital readout, as well as a much larger pixel matrix, potentially affecting the pixel electric field and charge sharing.

The paper is organized as follows: section 2 describes the RD50-MPW4; section 3 describes the TPA-TCT experiment setup; section 4 shows the experiment results and section 5 closes with conclusions.

\section{The RD50-MPW4 chip}
The RD50-MPW4 is a DMAPS chip manufactured in the 150 nm HV-CMOS process from LFoundry, in a substrate with a resistivity of 1.9 k\(\Omega\cdot\)cm and 280 \(\upmu\)m of thickness.

\begin{figure}[t]
	\centering
    \includegraphics[width=\columnwidth]{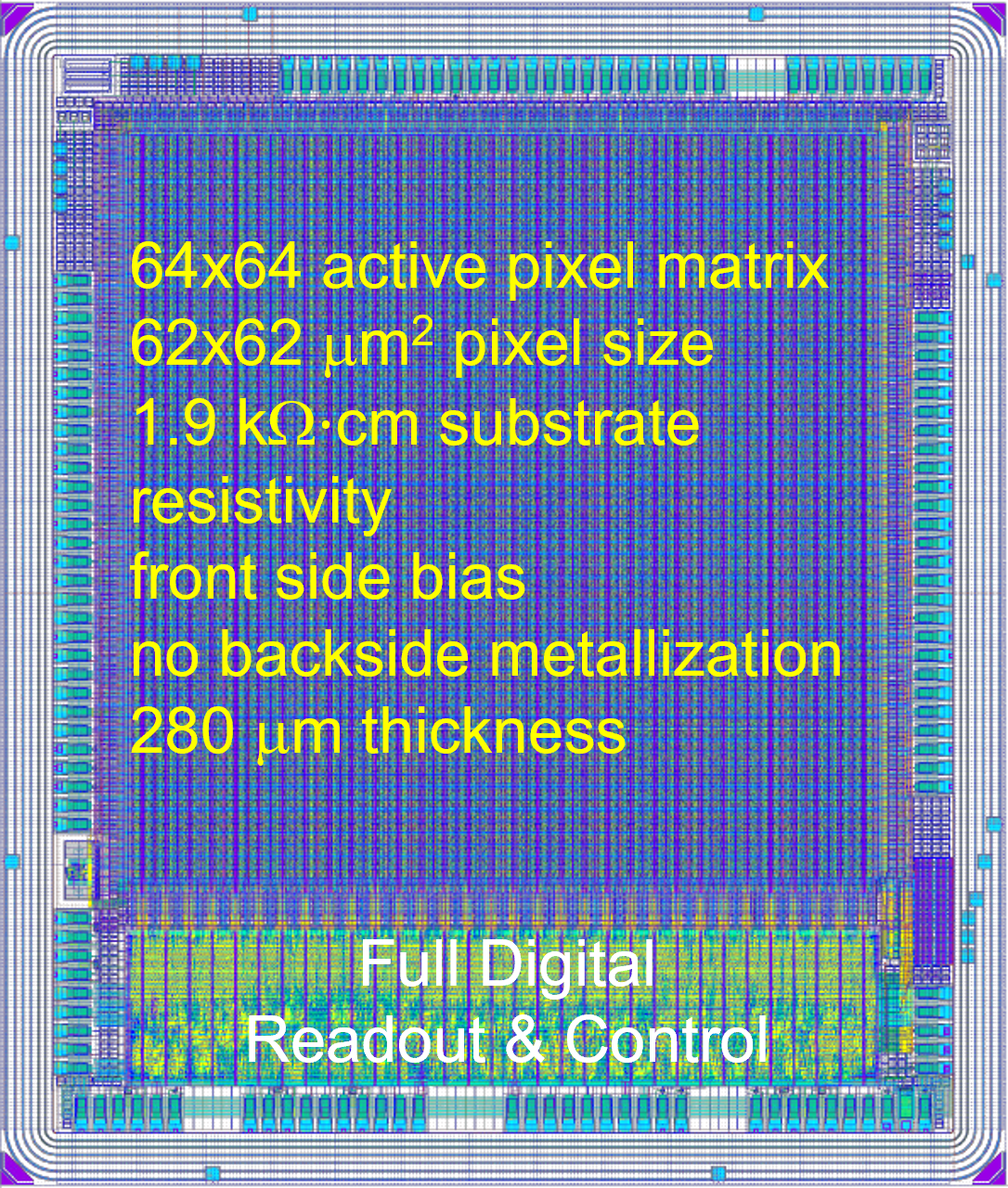}
	\caption{Layout view of the RD50-MPW4 DMAPS chip.}
    \label{MPW4_chip}
\end{figure}

The layout view is shown in Fig. \ref{MPW4_chip}, with the 64\(\times\)64 pixel detector matrix on top and the digital periphery at the bottom.

\begin{figure}[t]
	\centering
    \includegraphics[width=\columnwidth]{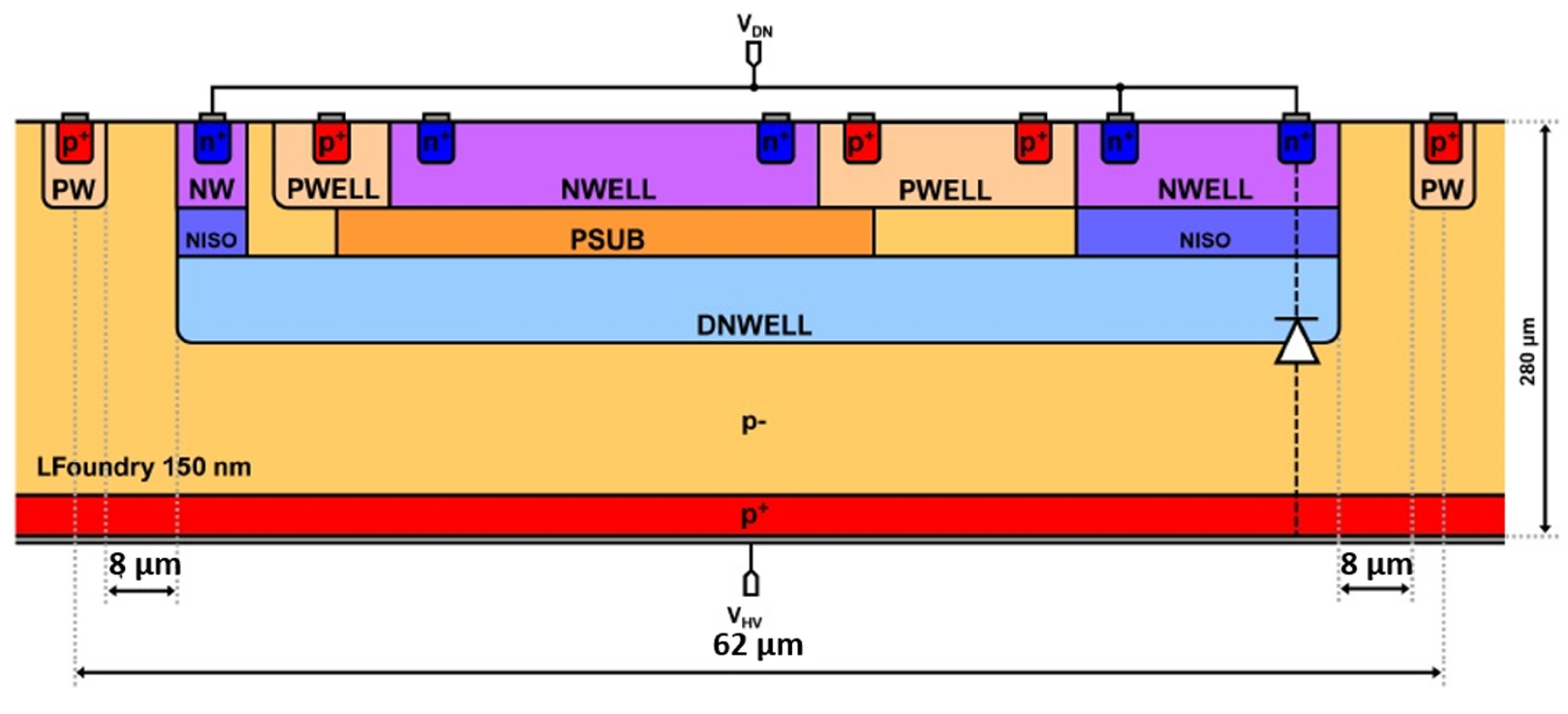}
	\caption{Cross section of an RD50-MPW4 pixel detector.}
    \label{pixel_crossection}
\end{figure}

Each pixel has a size of 62 \(\upmu\)m~\(\times\)~62 \(\upmu\)m. Every pixel is an embedded detector formed by a deep n-well (DNWELL) as collecting electrode and the high resistivity p-substrate as seen in Fig. \ref{pixel_crossection}. The DNWELL is biased at 1.75 V by a stack of two n-type layers (n-well and n-type isolated niso implant). The p-substrate is biased at reverse voltage to create the depletion region. A ring of a floating p$^+$ diffusion, separated by 8 $\upmu$m from the DNWELL, surrounds the entire pixel as a p-stop to avoid charge trapping in the surrounding Shallow Trench Isolation that separates one pixel from the neighbors.  

\begin{figure}[t]
	\centering
    \includegraphics[width=\columnwidth]{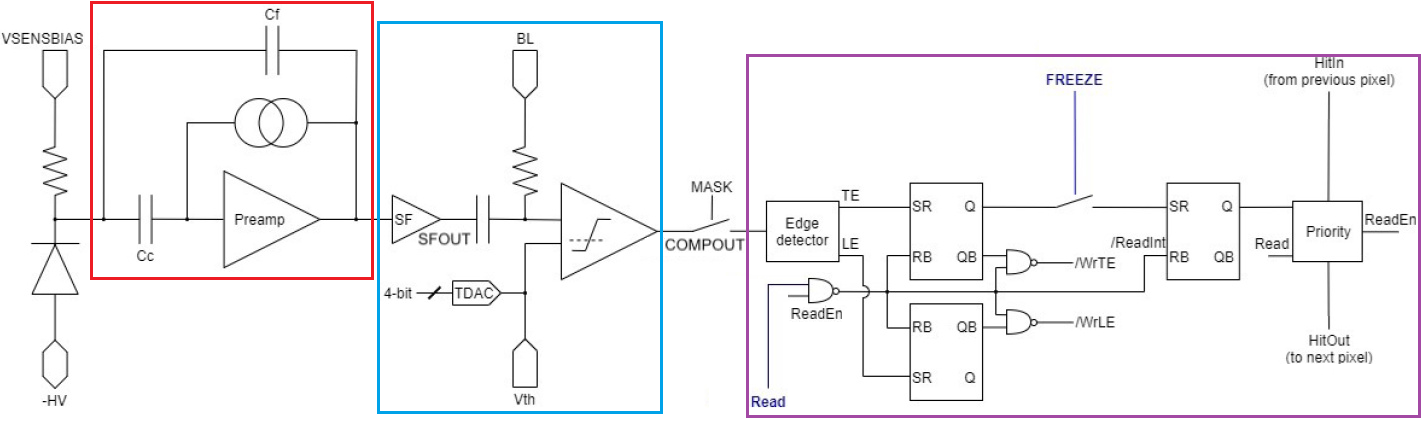}
	\caption{Schematic of the RD50-MPW4 DMAPS pixel electronics. Analog front-end: Charge sensitive amplifier in the red square, discriminator in the blue square. Digital back-end in the purple square.}
    \label{pixel_frontend}
\end{figure}

Figure \ref{pixel_frontend} shows the architecture of the pixel electronics. The analog front end electronics comprises a biasing circuit for the detector sensor, a pre-amplifier (a Charge Sensitive Amplifier, CSA), a filter and a discriminator (comparator plus DAC adjustable threshold). The discriminator output is processed by a digital backend. 

The detector sensor is negatively biased through the anode (top-side), so the coupling to the CSA is capacitive to have the input transistor of the amplifier at a different DC potential than the n-well of the sensor.

A source follower (SF) buffer produces the SFOut signal. That SFOut signal, if the particular pixel is selected, is sent to an analog buffer to make the signal available in an external pad for readout with an oscilloscope.

The comparator threshold is adjustable by a 4-bit DAC. If the signal amplitude is over the threshold, the comparator generates a COMPOUT signal that is temporarily stored in the digital backend as a valid detection. The in-pixel digital electronics generates a short pulse, the HitBus signal, when COMPOUT flips high, indicating the detection of the leading edge of COMPOUT. For the selected pixel, the HitBus signal is also externally available for analog monitoring. 

All pixels receive a global 8-bit Gray-encoded timestamp running at 40 MHz. When a hit is detected, the event timestamps are recorded. The digital stage includes two 8-bit RAMs for storing the time stamp of rising and trailing edges of the discriminator and another 8-bit ROM memory to determine the address of the pixel. 

The 64 pixel columns, organized in double columns, terminate in an End Of Column (EOC) bottom block, responsible for pixel configuration, readout of the particle hits in pixels and temporary storage of the data read from its double column. The EOC is handled by a control unit (CU), which also puts the retrieved information in a transmission FIFO (First In, First Out) memory. The CU manages the commands stored in the Control Status Registers (CSRs), all of them interconnected with a Wishbone bus. This is a a general purpose System On Chip interface between Intellectual Properties cores that defines standard data exchange, supported by the OpenCores organization \cite{Wishbone-2012}.

The RD50-MPW4 digital periphery implements digital readout and configuration. The control registers are divided in two categories: End Of Column (EOC) and peripheral. The EOC control registers are responsible for the configuration bits of the pixels in a double column. There is one EOC per double pixel column, which receives the hit data from pixels with higher addresses first and stores it in a 24-bit FIFO with a capacity for 16 hits. The peripheral registers enable or disable data transmission, configure the transmission protocol, set up the timestamp counter and DACs for the bias block. The control unit is a finite automaton responsible for handling the data flow from the EOC FIFOs to the transmission FIFO (TX FIFO). A Data Transmission Unit is responsible for reading the TX FIFO, packing and serializing the frames. Data is sent out at a speed of 640 MHz via an LVDS serial link.

For data acquisition and control, the RD50-MPW4 uses the Caribou system. It is a versatile data acquisition platform developed for the characterization and testing of silicon pixel detector prototypes in collaborative research projects \cite{Caribou-2025}.

\begin{figure}[t]
	\centering
    \includegraphics[width=\columnwidth]{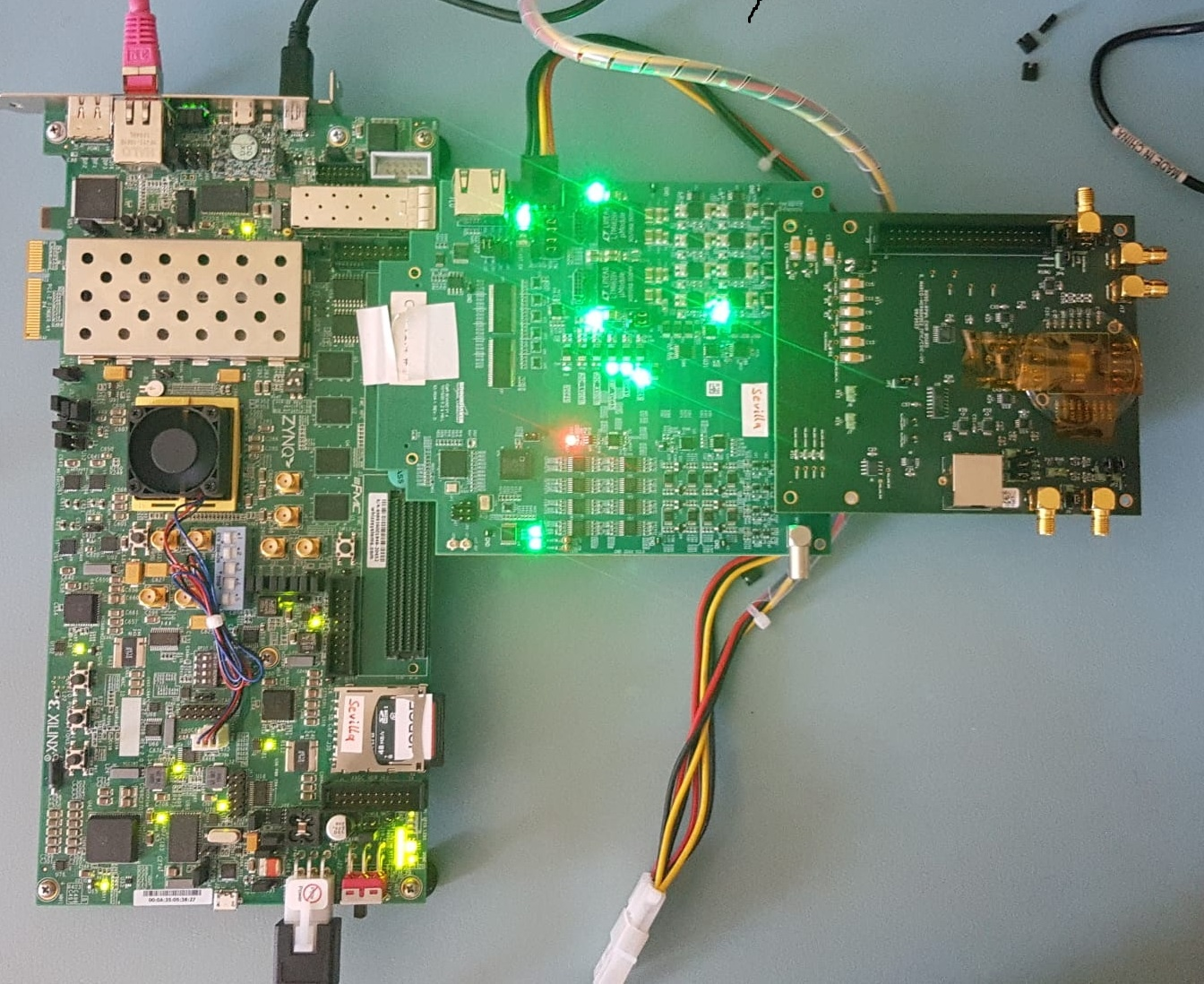}
	\caption{Data acquisition hardware setup based on the Caribou system, including the detector board hosting the RD50-MPW4 (full right), CaR Board (center, brilliant green LEDs), and Zynq ZC706 FPGA (full left, pale green LEDs).}
    \label{DAQ_caribou}
\end{figure}

The Caribou hardware setup comprises a detector board hosting the RD50-MPW4 chip and a signal conditioning card (CaR), which provides the required biasing and auxiliary services. The detector board is connected to the CaR through an FMC connector, while the CaR features an additional FMC interface used to connect it to a commercial Xilinx Zynq ZC706 board incorporating a Zynq-7000 FPGA system-on-chip (FPGA SoC). The data acquisition hardware setup is shown in Fig. \ref{DAQ_caribou}.

\begin{figure}[t]
	\centering
    \includegraphics[width=\columnwidth]{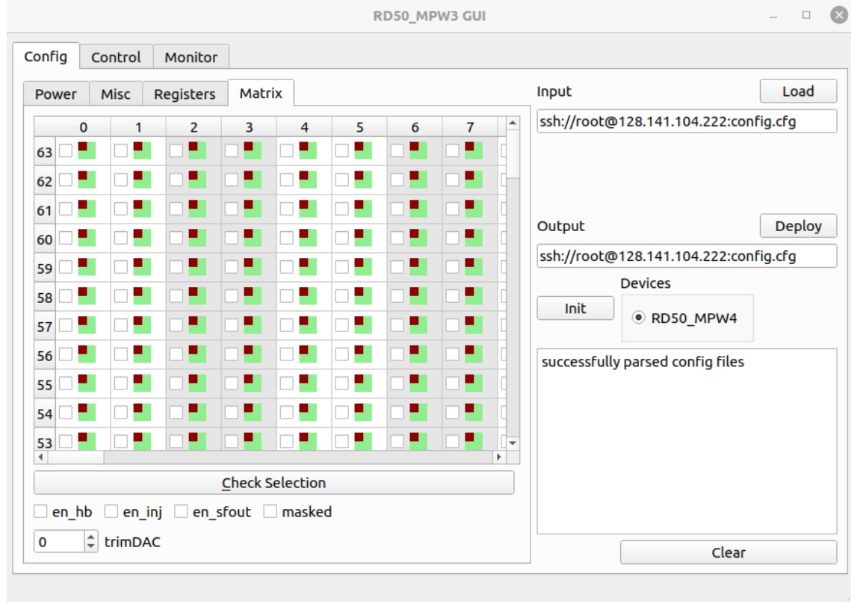}
	\caption{Graphical User Interface for Peary DAQ software framework.}
    \label{GUI_peary}
\end{figure}

The FPGA SoC board is connected to the control laptop via Ethernet using a TCP/IP, enabling remote control and data exchange with the RD50-MPW4 chip. The system is operated using the Peary DAQ software framework \cite{Peary-2025}, which runs on a Linux operating system executed on the Zynq-7000 SoC. The Peary framework is used to control the FPGA logic that communicates with the RD50-MPW4. An associated graphical user interface (GUI) provided within the Peary framework is used to facilitate remote operation of the system. The GUI is shown in Fig. \ref{GUI_peary}.

The implemented commands include powering up/down, resetting, configuring, calibrating, adjusting the discriminator threshold, choosing pixels to mask or to read, routing a pixel SFOUT signal to the driving buffer for analog monitoring, making pixel data being read out from the FIFO, registering the amount of hits along a time frame, recording ToT (Time over Threshold) spectrums, injecting electrical test pulses and masking noisy pixels if any. 

\section{TPA-TCT Setup}
\begin{figure}[t]
	\centering
	\begin{subfigure}[t]{\columnwidth}
		\centering
		\includegraphics[width=\columnwidth]{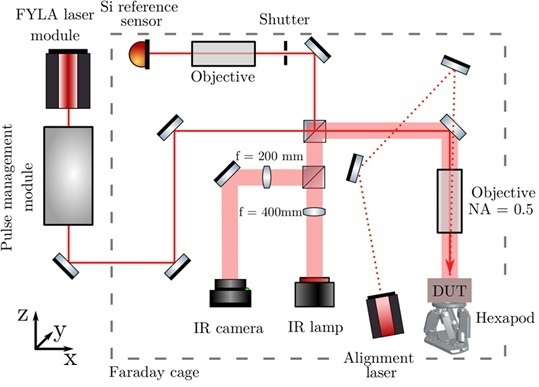}
		\caption{}
		\label{TPA_setup_a}
	\end{subfigure}

	\vspace{0.5em}

	\begin{subfigure}[t]{\columnwidth}
		\centering
		\includegraphics[width=\columnwidth]{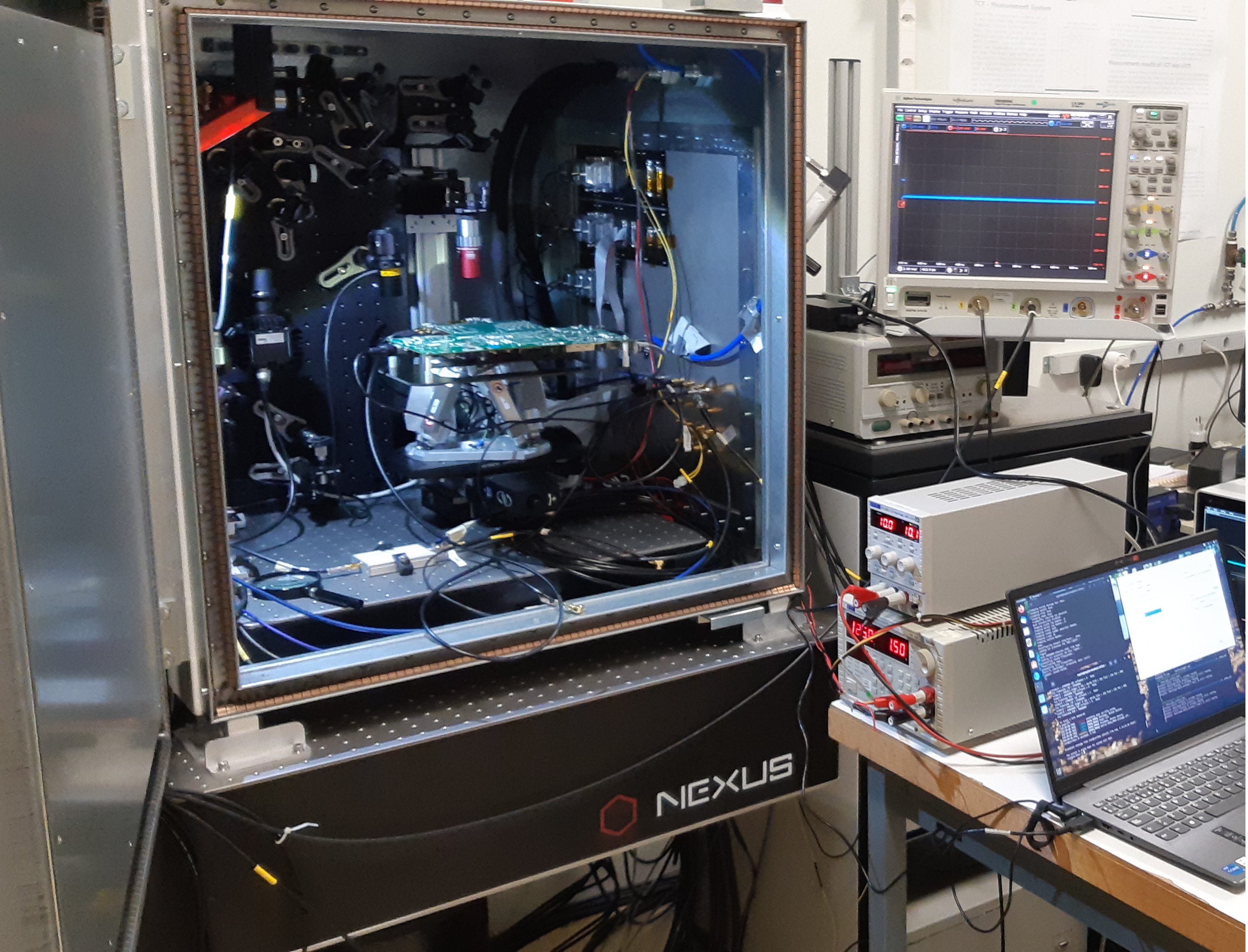}
		\caption{}
		\label{TPA_setup_b}
	\end{subfigure}

	\caption{TPA setup: (a) top, schematic of the TPA setup and (b) bottom, photograph of the real experimental setup.}
	\label{TPA_setup}
\end{figure}

The RD50-MPW4 pixel detector was characterized using the TPA-TCT measurement setup in the CERN SSD laboratory \cite{TPA-2021}. The general TPA-TCT laser setup is shown in Fig. \ref{TPA_setup}.

The laser setup employs an LFC1500X laser from the company FYLA, \cite{Laser-2022}, with a 1550 nm wavelength and a pulse duration of 400 fs. The laser light propagates in free-space optics up to an input port inside a Holland EMI (Electro Magnetic Interference) shielded measurement box. Inside the EMI box, light is driven by mirrors up to a focusing objective with numerical aperture (NA) of 0.5.  The laser pulse repetition rate typically used can vary from 1 kHz to 100 kHz and can be further reduced to 1~pulse\,s$^{-1}$ by means of a magnetoacoustic pulse picker situated along the free-space optical path. A local photodetector near the final target generates the trigger to capture any signal coming from the pixel detector. 

A carbon-fiber tray supports the electronic setup, corresponding to the device under test (DUT), and is bolted to a mobile hexapod platform with 100 nm positioning precision.

\begin{figure}[t]
	\centering
    \includegraphics[width=\columnwidth]{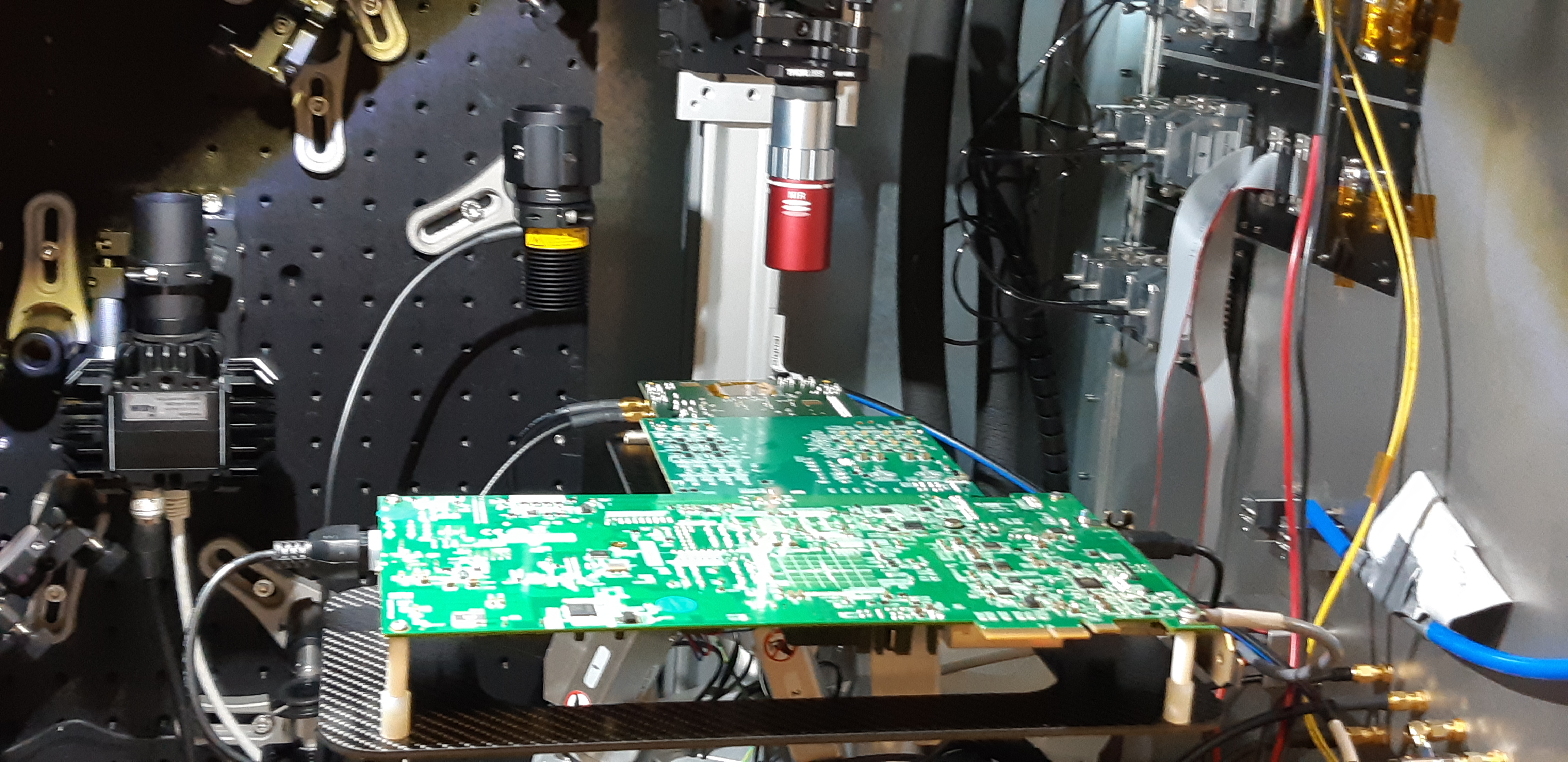}
	\caption{DUT in the hexapod platform.}
    \label{DUT_hexapod}
\end{figure}

Figure \ref{DUT_hexapod} shows the electronics setup under the laser optics, which consists of the detector board hosting the RD50-MPW4, the CaR board, and the Zynq ZC706 FPGA board. These boards are interconnected in cascade through FMC connectors and are all mounted on the tray attached to the hexapod platform. All cards are oriented face down, allowing laser light to pass through a window in the RD50-MPW4 detector board.

Right under the optical objective is the RD50-MPW4 detector board, face down, with the laser light passing through the window cut in the PCB (Printed Circuit Board). This optical objective is focused in a fixed point. This means that to obtain a focused position in (X,Y,Z), the hexapod is used to move the entire DUT setup.

The laser pulses impact the chip from the backside, where there is no metallization reflecting back the light. As in our setup the laser wavelength is in the C band (1550 nm), silicon is transparent if the intensity is low enough so there is no need of polishing the chip back side. The setup is described in full detail in \cite{TPA-2021}.

For this experiment, a pulse energy of 250 pJ is selected, a value sufficiently low to avoid generating electron–hole plasmas in the silicon pixel detector \cite{Laser-2022,Plasma-2021}. The ambient temperature is controlled at 25$^\circ$C. Signals were recorded in real time using an oscilloscope. Pixel analog signals from SFOut are the primary detection signal that was collected, but digital signals from the Hitbus were collected as well to ensure that the analog signal was generated by the impacting laser pulse and was not a noise dark count.	
We issued 1000 pulses by second and record in the scope the signal coming from the pixel detector. To reduce noise in the displayed waveforms, the oscilloscope traces were set to show the average of 128 consecutive pulses. An 128 samples average is enough to get clear traces to read the maximum amplitude, rise time and time over threshold without noisy spikes. 

The (X,Y) beam position is read from the hexapod stage. Initially, the light is focused to a point just on the DUT surface (backside of the RD50-MPW4 chip). The focal point within the DUT is set by moving the hexapod along the Z axis, with the stage coordinate corrected for the silicon refractive index according to equations \eqref{eq1:refractionindexcorrection}, \eqref{eq2:refractionindexcorrection}, and \eqref{eq3:refractionindexcorrection}. A detailed explanation is provided in \cite{TPA-2021}.

\begin{eqnarray}
	z_{\rm Si} &=& z_{\rm stage} \sqrt{\frac{z_R \pi n_{\rm Si}^3}{z_R \pi n_{\rm Si} - \lambda n_{\rm Si}^2 + \lambda}} 
	\label{eq1:refractionindexcorrection} \\
	w_0 &=& \frac{\lambda}{\pi \theta} \approx \frac{\lambda}{\pi \mathit{NA}} 
	\label{eq2:refractionindexcorrection} \\
	z_R &=& \frac{\pi w_0^2 n_{\rm Si}}{\lambda} 
	\label{eq3:refractionindexcorrection}
\end{eqnarray}

The hexapod displacement along the Z-axis is named as $z_{\rm stage}$, whereas the real focal point position in silicon is defined as $z_{\rm Si}$. As the beam has a Gaussian shape \cite{TPA-2021}, $z_R$ is the Rayleigh length in silicon, $n_{Si}$ is the index of refraction of silicon, $\lambda$ is the laser wavelength, $w_0$ is the beam radius at the waist (focal point), $\theta$ is the divergence angle, and $\mathit{NA}=n_{Si} \cdot \sin\theta$ is the numerical aperture. 

The measured values \cite{Palomo-2025} were \mbox{$w_0=1.3$\,\textmu m} and \mbox{$z_R=10.4$\,\textmu m}, so the correction factor is \mbox{$z_{Si}= 3.77\cdot\,z_{stage}$}. 

The volume of charge generation is approximated as an ellipsoid with elongation $z_R$ and circular cross-section of radius $w_0$ \eqref{eq4:volumeellipsoid}, because most of the charge is generated within the first Rayleigh length \cite{SebastianThesis-2023}. The measured voxel volume was \mbox{$V_{\rm TPA}\approx 74$\,\textmu m$^3$}.

\begin{equation}
	V_{\rm TPA}=\frac{4}{3}\pi w_0^2 z_R
	\label{eq4:volumeellipsoid}
\end{equation}

\begin{figure}[t]
	\centering
	\begin{subfigure}[t]{0.48\columnwidth}
		\centering
		\includegraphics[width=\linewidth]{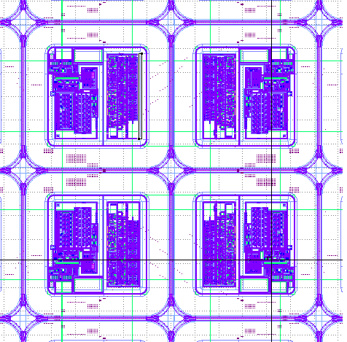}
		\caption{}
		\label{NIR_target_a}
	\end{subfigure}
	\hfill
	\begin{subfigure}[t]{0.48\columnwidth}
		\centering
		\includegraphics[width=\linewidth]{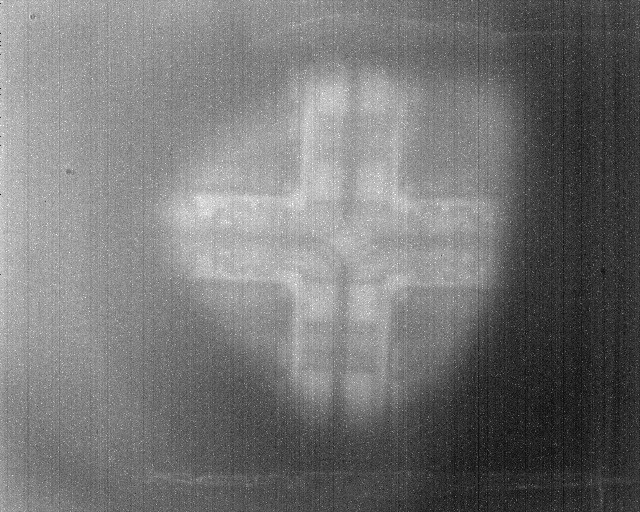}
		\caption{}
		\label{NIR_target_b}
	\end{subfigure}

	\vspace{0.8em}

	\begin{subfigure}[t]{0.48\columnwidth}
		\centering
		\includegraphics[width=\linewidth]{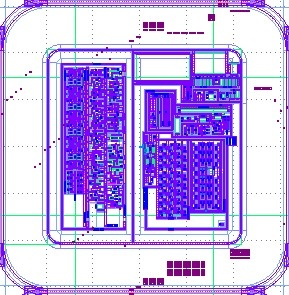}
		\caption{}
		\label{NIR_target_c}
	\end{subfigure}
	\hfill
	\begin{subfigure}[t]{0.48\columnwidth}
		\centering
		\includegraphics[width=\linewidth]{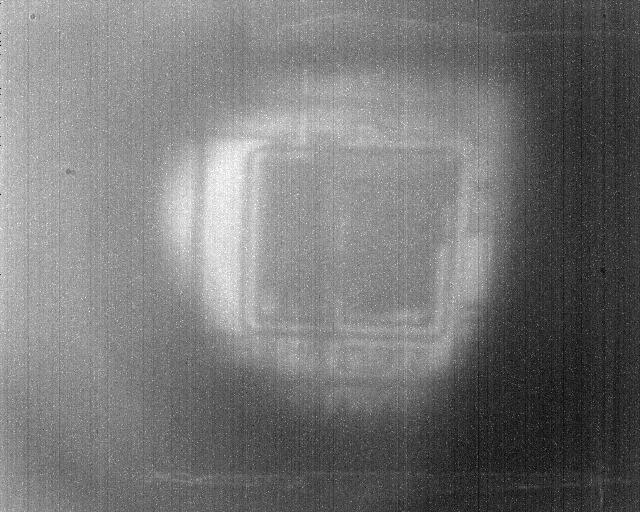}
		\caption{}
		\label{NIR_target_d}
	\end{subfigure}

	\caption{NIR images for coarse target localization: (a) layout image of a four-pixel group, (b) backside NIR image of the center (cross) of a four-pixel group, (c) detailed layout view of a single pixel and (d) backside NIR image of a single pixel.}
	\label{NIR_target}
\end{figure}

The laser facility has a Near Infrared (NIR) camera used for a coarse location of the target and tray alignment. The NIR camera coarse localization images are shown in the Fig. \ref{NIR_target}. Images are acquired using NIR illumination, to which silicon is transparent, allowing the metallization of the monolithic pixel detector to be clearly observed.

Tray alignment is particularly critical in our case because the electronics setup is on its own tray, bolted to the hexapod. Even with a carbon fiber composite tray, thermal stresses generate misalignments in the $\upmu$m range. 

TPA measurements need a precise sample alignment to get an orthogonal light beam incident to the sample surface. Because of the small pixel size of the MPW4, we corrected the tilt using the microscope image. By moving the sample in Z for a given X,Y position until the image is sharp, we can obtain the location of the surface in XYZ. Repeating this step for several X,Y allows us to perform the fit and the tilt correction. This optical tilting correction is not dependent on any charge measurement by any MPW4 pixel so we can extend the range in XY across the whole chip. We measured tilt angles of u= -2.907$^\circ$, v= 2.018$^\circ$, which were then corrected for in all following measurements.

The digital logic in the digital periphery of the RD50-MPW4 delivers hitmap data. A map of hits gives the correlation between the laser focal point XY position and any specific pixel, even counting the number of hits of any particular pixel. The first use of the hitmap feature was to precisely identify one central pixel by line XY scanning along the chip (movement made by the hexapod platform). Digital logic also enables the discriminator threshold at every pixel to be adjusted to avoid noise hits.
	
After coordinate calibration, the entire pixel matrix was explored in a zig-zag manner, with the focal point at 113 $\mu$m depth, selecting every pixel and recording its signal response. Since every pixel detector in the chip was functional, a center pixel was selected to further analyze it. The center location at the 64$\times$64 detector matrix, pixel (33,31), was also useful to avoid any uncorrected tilting. 

\section{Results}
The RD50-MPW4 pixel detectors were biased at -90 V, with the discriminator threshold externally set at 0.94 V (internally, the electronics transform that into a real discriminator threshold of 40 mV). For the top-side biased MPW4, the breakdown happens at -190 V$_{\mathrm{bias}}$, with a bias safe range up to -120 V at 10$^{\circ}$C \cite{Vilella-2025}. As the MPW4 samples available are limited and the room temperature was 25$^{\circ}$C, the sensor was operated at a conservative -90 V$_{\mathrm{bias}}$ to reduce leakage current and to prevent damage. In all cases, the laser light traveled from the backside to the front side of the chip (backside illumination). Each measurement was taken to be an average of a total of 128 signal amplitudes. 

The TPA mapping of detectors has a precision determined by the beam radius at the focal point $w_0$=1.3 $\mu$m and the Raleigh length in silicon $z_R$=10.4 $\mu$m, as discussed in the previous section. In XY maps that means a maximum resolution of w$_0$; for XZ or YZ scans it means a maximum vertical resolution of $\sim$5$~\mu$m. Once the pulse energy is larger than the TPA threshold, the energy pulse resolution is not critical as we only need to record signals over the noise floor. Here the laser pulse energy is around 1 pJ. The real limitation is the upper limit for the pulse energy as eh plasma formation has to be avoided. Plasma formation is easily recognized in the increase of signal return to the baseline, see \cite{Plasma-2021}, making it harder to determine the signal amplitude. For that reason, a pulse energy of 250$\pm$1 pJ was chosen during the initial pixel exploration, as described in previous section.

\subsection{Hit Detection Efficiency}
The first measurement was that of the Hit Detection Efficiency (HDE). With the pulse energy at 250 pJ, the laser was focused at the center of pixel (33,31) at a focal depth of 100 $\upmu$m in the middle of this pixel. The pulse repetition rate was adjusted to 1 kHz and hits were accumulated for a full second. Exactly 1000 hits were detected in the pixel (33,31), which validates that the pixel detector has a HDE of 100\% under the specified conditions. The relevance of the HDE measurement comes from the necessity to determine a laser pulse repetition rate. If it is too low, the entire experiment will last too long, but if it is too high, pile up could appear. At 1 kHz, both problems are avoided.

\subsection{Z-scan}
\begin{figure}[t]
	\centering
	\begin{subfigure}[t]{\columnwidth}
		\centering
		\includegraphics[width=\linewidth]{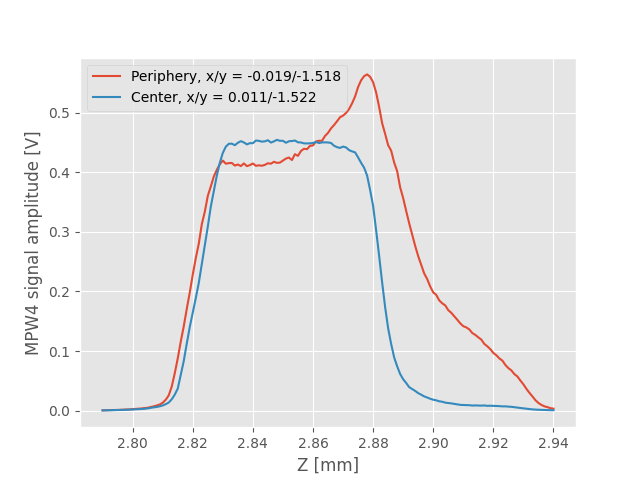}
		\caption{}
		\label{Z_Scans_a}
	\end{subfigure}
    
	\begin{subfigure}[t]{0.5\columnwidth}
		\centering
		\includegraphics[width=\linewidth]{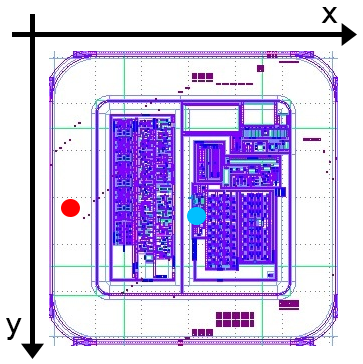}
		\caption{}
		\label{Z_Scans_b}
	\end{subfigure}

	\caption{Initial Z-scan calibration: (a) signal amplitude as a function of depth for positions at the pixel center (blue) and near the pixel edge (red); (b) corresponding (X,Y) positions on the chip.}
	\label{Z_Scans}
\end{figure}

The second measurement consisted of a Z-scan calibration: Two fixed (X,Y) positions within a single pixel were selected, and the laser focal point was scanned along the depth (Z) direction in steps of 1 $\upmu$m (refraction-index-corrected displacement inside silicon). At each depth position, the corresponding signal amplitude was recorded. The results are shown in Fig. \ref{Z_Scans}. The blue curve corresponds to an (X,Y) position approximately in the pixel center, a place where chip metallization is sparse. The metallization on the chip front side, on top of the transistors, is very prominent. The red curve corresponds to an (X,Y) position near the pixel edge, where the chip metallization is dense: when the focal point is far from the chip top side, the red curve shows less signal amplitude because it is in the edge of the detector active volume; when the focal point is approaching the chip top side, the red curve shows more signal amplitude up to a peak resulting from light reflected back from the metallization. This particular test is very important for weighting the signal amplitudes in the subsequent tests to compensate the effect of metallization reflections. 

From the Z-scan it can be concluded that the mean signal response amplitude is $\sim$450 mV, with bigger amplitudes (up to 530 mV) resulting from reflections from the top metallization, and lower amplitudes ($\sim$405 mV) when the focal point is near the position of the pixel guard rings.

\subsection{XZ-scan}
\begin{figure}[t]
	\centering
	\begin{subfigure}[t]{\columnwidth}
		\centering
		\includegraphics[width=\linewidth]{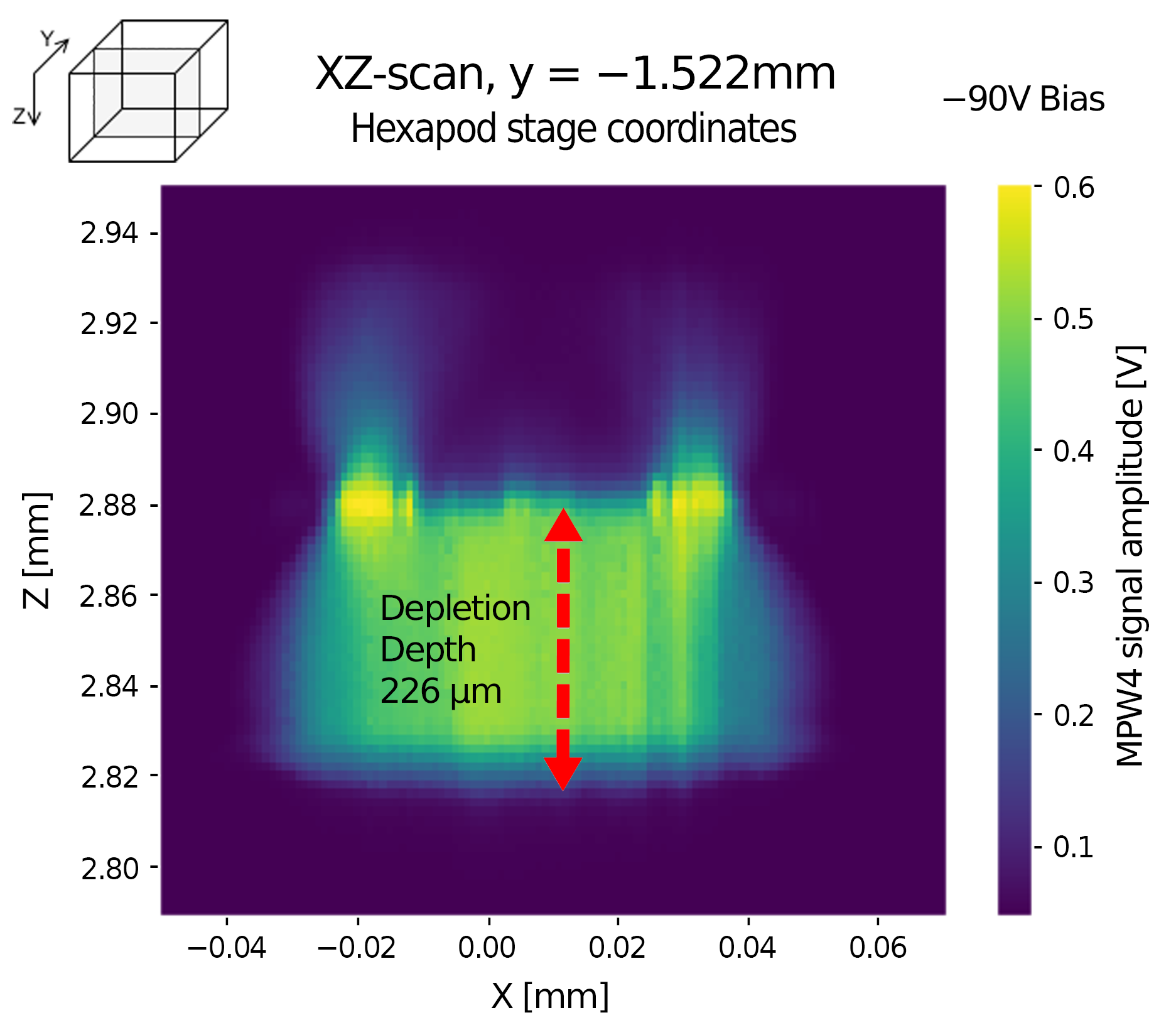}
		\caption{}
		\label{XZ_Scans_a}
	\end{subfigure}
    
	\begin{subfigure}[t]{0.5\columnwidth}
		\centering
		\includegraphics[width=\linewidth]{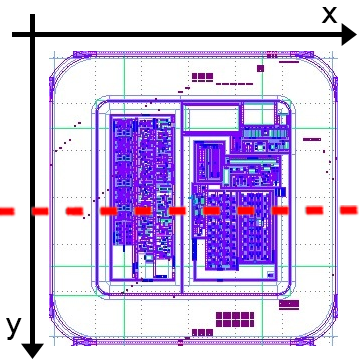}
		\caption{}
		\label{XZ_Scans_b}
	\end{subfigure}

	\caption{XZ scan at a fixed Y = 85 $\upmu$m, where X ranges from 0 to 62 $\upmu$m and Z from 0 to 264 $\upmu$m: (a) measured values across the electronic structure and (b) corresponding positions on the chip.}
	\label{XZ_Scans}
\end{figure}

The subsequent measurement consisted of an XZ-scan, as shown in Fig. \ref{XZ_Scans}. 

In this case, the laser focal point was scanned from the backside to the topside along the z-coordinate and also in a straight line along one Y direction in the middle of the chip (y=85 $\upmu$m). The image shows the amplitude of the recorded detector signal in correlation with the focal point position. After considering the optical refraction correction (equation (\ref{eq1:refractionindexcorrection})), an effective depletion depth W of 226$\pm$5 $\upmu$m was extracted for the detector. In fact, the full XZ-scan is a vertical section of the depletion volume. The term "effective depletion" is used because a non-zero response may also include diffusion from a non-depleted region.  The length uncertainty comes from the shape of the ellipsoid volume where two-photon absorption happens ($z_R$=10.4 $\mu$m). From the XZ scan, the bottom boundaries of the depletion volume were determined when the signal recorded had 45 mV of amplitude, a 10\% of the mean maximum recorded amplitude ($\approx$ 450 mV), as discussed in the previous subsection. Halo-like structures are observed in the image and are attributed to parasitic reflections, in particular from the metallization pixel top side, marking the edges of the detector. Given a total pixel thickness of 280 $\upmu$m, the measured depletion depth indicates that, at a bias voltage of -90 V, the detector is near to full depletion (226$~\mu$m/280$~\mu$m$~\simeq~$81\%).

\subsection{XY-tomography}

\begin{figure}[!htbp]
	\centering
	\begin{subfigure}[t]{0.88\columnwidth}
		\centering
		\includegraphics[width=\columnwidth]{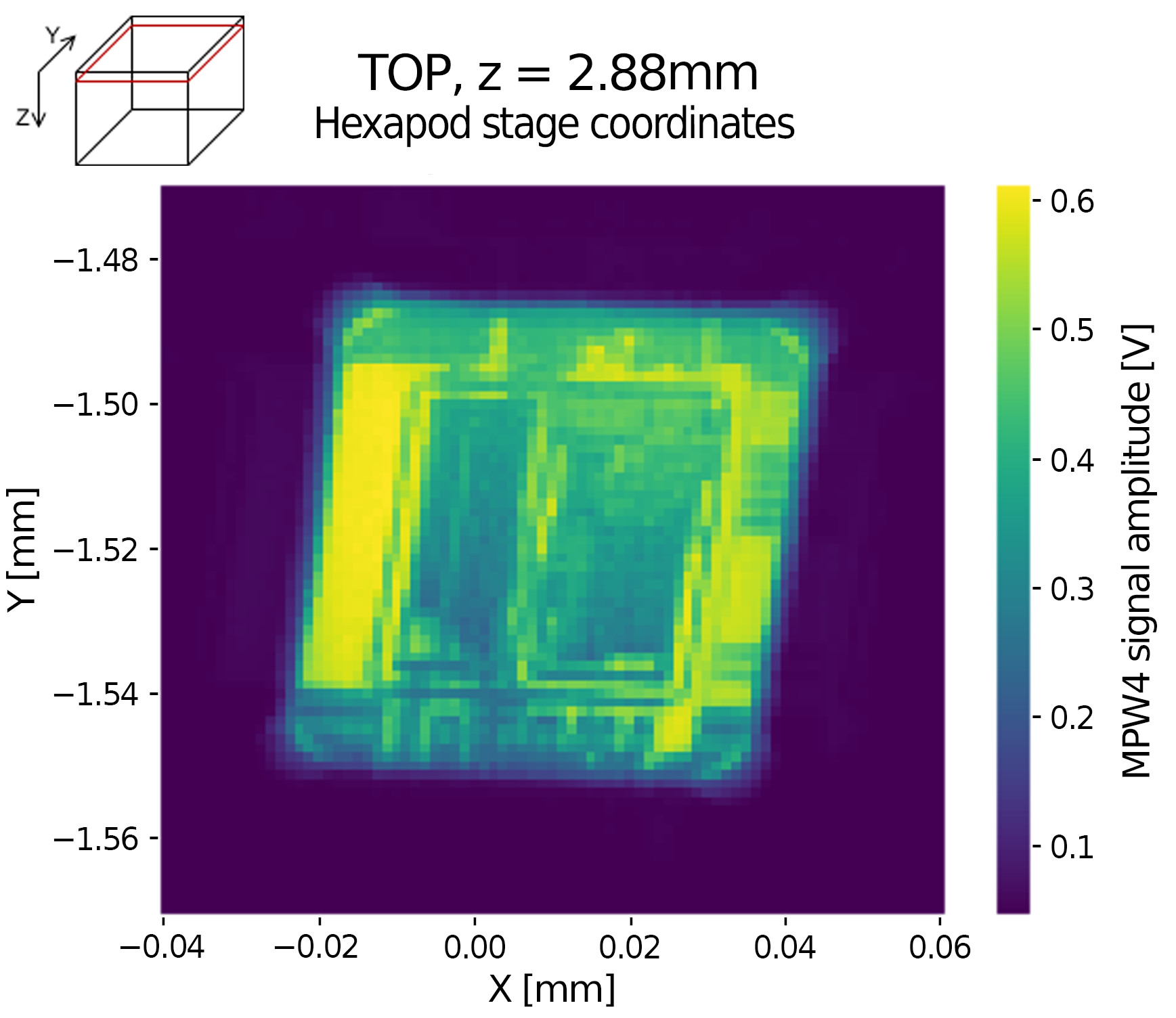}
		\caption{}
		\label{XY_Tomography_Top}
	\end{subfigure}

	\begin{subfigure}[t]{0.88\columnwidth}
		\centering
		\includegraphics[width=\columnwidth]{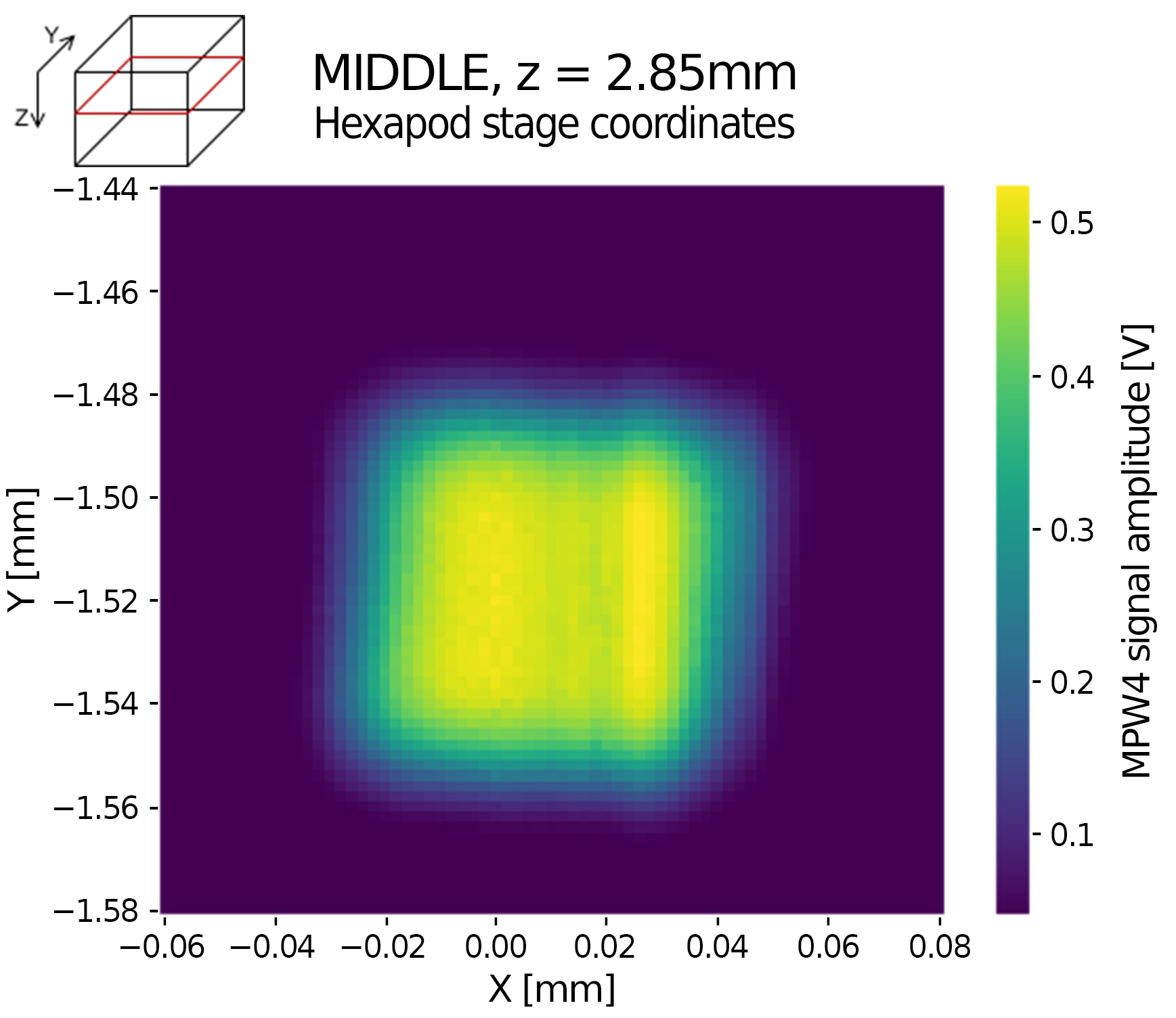}
		\caption{}
		\label{XY_Tomography_Middle}
	\end{subfigure}

	\begin{subfigure}[t]{0.88\columnwidth}
		\centering
		\includegraphics[width=\columnwidth]{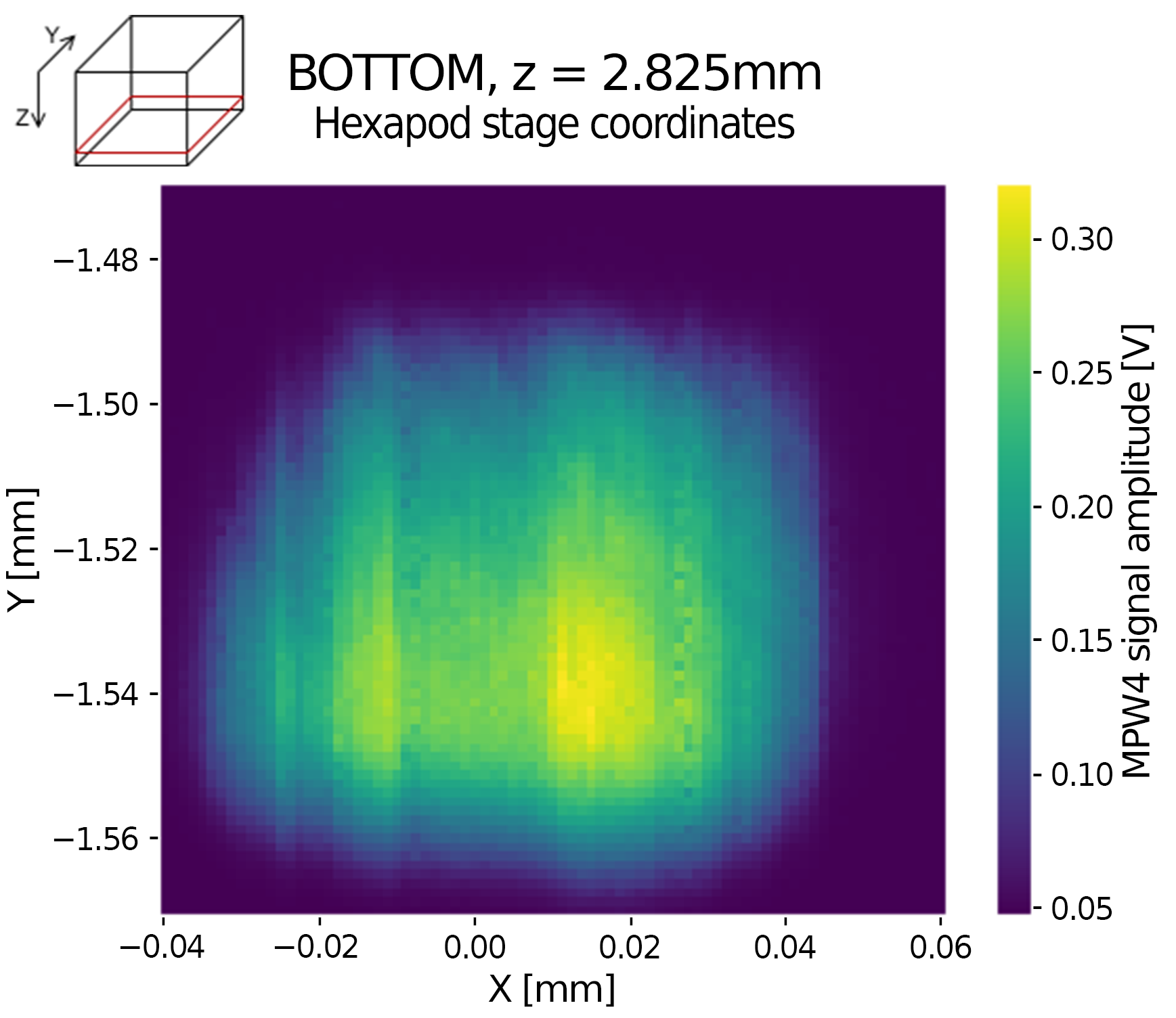}
		\caption{}
		\label{XY_Tomography_Bottom}
	\end{subfigure}
    
	\caption{XY-scan pixel tomography: (a) XY tomography near the top of the pixel at a depth of 3 $\upmu$m from the top; (b) XY tomography near the middle of the pixel at a depth of 113 $\upmu$m from the top; (c) XY tomography near the bottom of the pixel at a depth of 245 $\upmu$m from the top.}
	\label{XY_Tomography}
\end{figure}

The next measurement was an XY-scan pixel tomography. A total of three pixel sensor slices are shown in Fig. \ref{XY_Tomography_Top}, Fig. \ref{XY_Tomography_Middle}, and Fig. \ref{XY_Tomography_Bottom}. The XY-scan covers the entire sensor area and every slice has the focal point at a different depth.  

Figure \ref{XY_Tomography_Top} has the focal point depth at 3 $\upmu$m under the transistors. As the focal point has a beam waist radius of $w_0$=1.3$\mu$m, that is the minimum spot size in the maps. The signal amplitude is reduced, as expected, in the areas with a higher density of microelectronics. The signal amplitude is increased in the areas where the metallization is abundant because of the reflections. The image tilting is a consequence of a tilting correction made with the hexapod to compensate the small thermal bending of the entire electronics setup.

The subsequent measurement was performed with the laser focal point positioned at a depth of 113 $\upmu$m, corresponding approximately to the center of the effective depletion region, as shown in Fig. \ref{XY_Tomography_Middle}. As expected, the signal amplitude is high in the center, with a very homogeneous amplitude distribution and diminishing toward the edges, showing the edges of the pn-junction electric field (as there is practically no signal if there is no E-field). At this depth there is still a slight effect resulting from reflections from the top.

The last slice was made with the focal point at 245 $\upmu$m depth, a little beyond in the limit of the depletion volume, as shown in Fig. \ref{XY_Tomography_Bottom}. The signal amplitude is now reduced and the edges are blurred.

As shown in Fig. \ref{pixel_crossection}, the detector sensor (or diode) at every pixel is formed by its DNWELL, p- substrate and p+ ohmic layer. From the XY maps it is clear that the active volume does not have the simple geometry of a rectangular prism centered at the DNWELL. The prism boundaries are well defined by the guard rings limits near the electronics layer. The boundaries start to blur at mid depth and instead, the prism boundaries blur at the end of the effective depletion, expanding beyond the guard rings XY position. The active volume's geometry strongly suggests that charge sharing will occur between neighboring pixels. 

\subsection{Charge sharing}
\begin{figure}[t]
	\centering
	\begin{subfigure}[t]{\columnwidth}
		\centering
		\includegraphics[width=\linewidth]{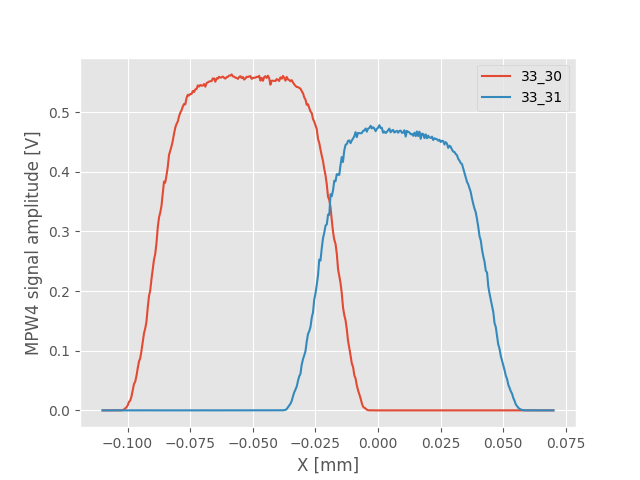}
		\caption{}
		\label{Charge_Sharing_a}
	\end{subfigure}
    
	\begin{subfigure}[t]{0.9\columnwidth}
		\centering
		\includegraphics[width=\linewidth]{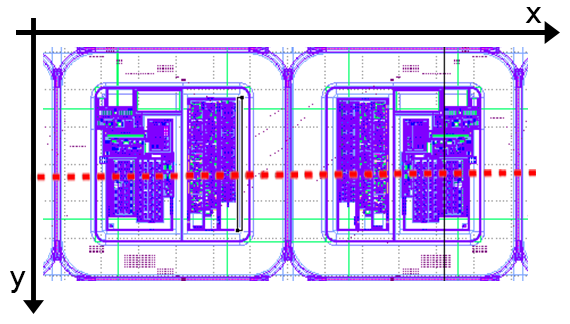}
		\caption{}
		\label{Charge_Sharing_b}
	\end{subfigure}

    \caption{Charge sharing measurement at a focal point of 113 $\upmu$m depth in the middle of the depletion volume and in the center of the pixel: (a) signal response along X, showing overlap between pixels (33,30) and (33,31), shown in red and blue, respectively; (b) corresponding (X,Y) positions on the chip.}
	\label{Charge_Sharing}
\end{figure}

The final measurement addressed charge-sharing effects. The laser focal point was positioned at a depth of 113 $\upmu$m, corresponding to the center of the depletion region, and laterally centered within the pixel in the XY plane, as shown in Fig. \ref{Charge_Sharing_b}. The focal point was moved along the X direction, starting from the left edge of pixel (33,30) up to the right edge of pixel (33,31). Both pixels' discriminator thresholds had the same value and same analog gain. The pixels were measured in separate scans.
	
A clear overlap of the signal response is observed in the inter-pixel region, as shown in Fig. \ref{Charge_Sharing_a}. Overlapping neighboring-pixel responses are consistent with charge sharing. In this region, a fraction of the charge generated near the periphery of pixel (33,30) is collected by the neighboring pixel (33,31). The signal amplitude is notably different in (33,30) than in (33,31), even at the same excitation conditions as the laser intensity is the same and with the same discriminator threshold. As every pixel is similar to any other one, the differences in amplitude response likely come from optical losses resulting from imperfections on the backside chip surface where the light beam goes through.

\section{Conclusion}
A laser wavelength suitable for two-photon absorption allows detailed mapping of the pixel sensitivity, providing a qualitative evaluation of the detector built-in electric field homogeneity. It is especially suitable for DMAPS because of the in-pixel electronics that becomes visible in the measurement. In this work we present the use of TPA to characterize a representative pixel of a DMAPS pixel matrix. Using a map in the XZ plane, it has been characterized the depletion depth of a pixel of the RD50-MPW4 HV CMOS sensor and found the 280~$\upmu$m sensor to be nearly fully depleted down to 226~$\upmu$m. In the XY plane, the hit detection efficiency was tested inside the pixel volume and found to be 100\% under the specified conditions. Overlapping neighboring-pixel responses were observed, consistent with charge sharing. Such measurements provide an effective approach for detector quality assessment prior to irradiation-induced degradation.

\section*{CRediT authorship contribution statement} 
\textbf{Francisco Rogelio Palomo:} Conceptualization, Methodology, Formal analysis, Software, Investigation, Visualization, Data Curation, Writing (original draft), Supervision, Project Administration. \textbf{Jorge Jiménez-Sánchez:} Conceptualization, Methodology, Formal analysis, Software, Investigation, Visualization, Data Curation, Writing (original draft). \textbf{Moritz Wiehe:} Resources, Software, Data curation, Validation. \textbf{Jory Sonneveld:} Investigation, Data curation, Validation, Writing (review \& editing). \textbf{Bernhard Pilsl:} Resources, Software, Validation. \textbf{Fernando Muñoz-Chavero:} Validation, Writing (review \& editing). \textbf{Raimon Casanova:} Resources, Validation. \textbf{Christian Irmler:} Resources, Validation. \textbf{Patrick Sieberer:} Resources, Validation. \textbf{Chenfan Zhang:} Resources, Validation. \textbf{Sinuo Zhang:} Resources, Validation. \textbf{Eva Vilella:} Resources, Funding acquisition, Writing (review \& editing), Supervision. \textbf{Michael Moll:} Resources, Funding acquisition, Writing (review \& editing), Supervision.

\section*{Declaration of Competing Interest} 
The authors declare that they have no known competing financial interests or personal relationships that could have appeared to influence the work reported in this paper.

\section*{Data availability} 
Data will be made available on request.

\section*{Acknowledgments} 
This work was supported by Grant PID2022-138510OB-C22 funded by MICIU/AEI/10.13039/501100011033 and, as appropriate, by “ERDF A way of making Europe”, and by the program “Contrato Predoctoral PIF del VI Plan Propio de Investigación y Transferencia, Convocatoria 2020” supported by the Universidad de Sevilla (Spain). This work has been performed in the framework of the CERN-RD50 collaboration.
 

\end{document}